\documentclass[11pt, oneside]{article}   	
\usepackage{geometry}                		
\usepackage{graphicx}				
\usepackage{natbib}
\usepackage{mnemonic}	   
\usepackage{amssymb}
\usepackage{color}
\usepackage{ulem}
\newcommand{\fracb}[2]{\left(\frac{#1}{#2}\right)}

\newcommand{\eqb}{\begin{equation}}
\newcommand{\eqe}{\end{equation}}

\definecolor{forestgreen}{rgb}{0.13, 0.55, 0.13}
\definecolor{green(html/cssgreen)}{rgb}{0.0, 0.5, 0.0}
\definecolor{orange}{rgb}{1.0, 0.5, 0.0}
\definecolor{Myorange}{cmyk}{0,0.42,1,0}
\definecolor{Myblue}{rgb}{0.10,0.15,1}
\definecolor{Mypurple}{rgb}{0.8,0.0,0.8}
\definecolor{Mygrey}{gray}{0.75}

\def \simless {\mathbin{\lower 3pt\hbox{$\rlap{\raise 5pt
              \hbox{$\char'074$}}\mathchar"7218$}}}
\def \simgreat {\mathbin{\lower 3pt\hbox{$\rlap{\raise 5pt
              \hbox{$\char'076$}}\mathchar"7218$}}}

\newcommand{\rL}{r_{_{\rm L}}}

\newcommand{\bh}{\beta_{_{\rm h}}}
\newcommand{\Gh}{\Gamma_{_{\rm h}}}

\newcommand{\Lj}{L_{_{\rm j}}}

\newcommand{\rrel}{R_{_{\rm rel}}}
\newcommand{\tb}{t_{_{\rm b}}}

\newcommand{\dash}{\textit{-}}
\newcommand{\bhbar}{\bar{\beta}_{_{\rm h}}}
\newcommand{\Ghbar}{\bar{\Gamma}_{_{\rm h}}}

\makeatletter
\newcommand{\do@openE}[1]{%
  \mbox{\fontsize{#1}\z@\cmuserif\symbol{"0190}}%
}
\newcommand{\openE}{\mathord{\mathchoice
  {\do@openE\tf@size}
  {\do@openE\tf@size}
  {\do@openE\sf@size}
  {\do@openE\ssf@size}
}}
\makeatother

\date{\today}

\begin{document}

\title{On the Composition of GRBs' Collapsar Jets}

\author{Omer Bromberg$^{1}$, Jonathan Granot$^2$,  Tsvi Piran$^3$}

\maketitle

\centerline{\footnotesize ${^1}${}Department of Astrophysical Sciences, Princeton University, 4 Ivy Ln., Princeton NJ 
08544, USA}
\centerline{\footnotesize${^2}${Department of Natural Sciences, The Open University of Israel, P.O.B 808,
Ra'anana 43537, Israel}}
\centerline{\footnotesize${^3}${Racah Institute of Physics, The Hebrew University, Jerusalem 91904, 
Israel}}

\begin{abstract}
The duration distribution of long Gamma Ray Bursts reveals a plateau at 
durations shorter than $\sim20$ s  (in the observer frame) and  a 
power-law decline at longer durations \citep{2012ApJ...749..110B}. 
Such a plateau arises naturally in the Collapsar model. In this model  
the engine has to operate  long enough to push the jet out of  the stellar envelope
and the observed duration of the burst is the difference between the 
engine's operation time and the jet breakout time. 
We compare the jet breakout time inferred from the duration
distribution ($\sim 10$ s in the burst's frame) to the {breakout time}
of a hydrodynamic jet ($\sim10$ s for typical parameters) and of a
Poynting flux dominated jet with the same overall energy ($\lesssim 1$
s). As only the former is compatible with the duration of the plateau
in the GRB duration distribution, we conclude that the jet is
hydrodynamic during most of the time that its head is within
the envelope of the progenitor star and around the time when
it emerges from the star. This would naturally arise if the
jet forms as a hydrodynamic jet in the first place or if it forms 
Poynting flux dominated but
dissipates most of its magnetic energy early on within the
progenitor star and emerges as a hydrodynamic jet.
\end{abstract}


\section{introduction}\label{sec:introduction}

Direct SNe-GRB observations as well as the location  of long GRBs (LGRBs) 
within star forming regions 
revealed that LGRBs arise during the death of massive stars 
\citep[see e.g.][for a review]{2006ARA&A..44..507W}. 
On the other hand the observed spectral and temporal properties show
that the prompt $\gamma$-rays are emitted within relativistic jets at
large distances ($\sim 10^{12}-10^{16}\;$cm) from the GRB progenitor
\citep[see e.g.][for a review]{2004RvMP...76.1143P}.
These observations are explained by the Collapsar model
\citep{1999ApJ...524..262M}.
According to this model\footnote{We adopt here
a broad definition of the Collapsar model, which involves any central
engine that launches a jet within a collapsing star, regardless of the
specific nature of the central engine or the composition of the jet.} a compact object that forms at the center of
the collapsing star, launches a jet that drills a hole through the
star. Once the jet breaks out from the stellar envelope it emits the
observed gamma-rays far away from the progenitor star.  
While the Collapsar model successfully addresses the question of how a
dying star produces a GRB, a major open question involves the nature
of the relativistic jet. In this work we {address this
question by examining} the implications of recent the understanding of
the propagation of hydrodynamic \citep[][hereafter
BNPS11]{2011ApJ...740..100B} and MHD \citep[][hereafter
BGLP14]{2014arXiv1402.4142B} jets within stellar envelopes.

Hydrodynamic jet propagation within stellar envelopes was studied both 
analytically \citep{2001PhRvL..87q1102M, 2003MNRAS.345..575M, 
2005ApJ...629..903L, 2011ApJ...740..100B} and
numerically \citep{1999ApJ...524..262M, 2000ApJ...531L.119A, 
2001ApJ...550..410M, 2003ApJ...586..356Z, 2005ApJ...629..903L,
2007ApJ...665..569M, 2009ApJ...699.1261M, 2013ApJ...777..162M},
while the propagation of a magnetic jet in stars was discussed in 
\citep{2003ApJ...599L...5P, 2007PhPl...14e6506U, 2009MNRAS.396.2038B, 
2013ApJ...764..148L, 2014arXiv1402.4142B,BromTchek14}.
These works show that
as long as the jet {does} not breach out of the star it
dissipates {most of the energy that reaches its head}.  
It follows that a minimal amount of energy is
needed to push the jet out of the star.  This is translates to a
minimal time, denoted the ``breakout time" {$\tb$}, that the
engine must operate for a successful breakout of the jet. If the
engine stops before this breakout time, the jet's head will not reach
the stellar surface, and a regular GRB won't arise leading to a
``failed jet" or a ``failed GRB".

A ``failed jet" might not go unnoticed.  {\it Low luminosity GRBs}
({\it ll}GRBs) are a distinct group of GRBs characterized by their low
(isotropic equivalent) luminosity, which is two orders of magnitude
lower than the luminosity of typical GRBs, as well as by their low
peak {photon} energy and their smooth, single peaked light
curves.  Because of their low luminosity {\it ll}GRBs are detected
only from nearby distances.  Even though just a few {\it ll}GRBs have
been observed, their overall rate {(per unit volume)} is
larger by about a factor of 10 than the overall rate of regular LGRBs
\citep{2006Natur.442.1014S,2006ApJ...645L.113C, 
2006Natur.442.1011P,2007ApJ...662.1111L,2007ApJ...657L..73G,
2011ApJ...726...32F}.
There have been numerous arguments 
suggesting that  {\it ll}GRBs  arise from ``failed jets" that don't  break out 
from the stellar envelope 
\citep{1998Natur.395..663K, 2001ApJ...550..410M,  2001ApJ...551..946T, 
2006Natur.442.1008C, 2007ApJ...664.1026W,
2007ApJ...667..351W, 2010ApJ...716..781K,2012ApJ...747...88N, 2011ApJ...739L..55B}.

After the jet emerges from the stellar envelope it dissipates {some of} its
energy at a large distance and produces the GRB.  The
overall behavior of the prompt emission does not vary significantly
during the burst (the second half of the prompt emission is rather
similar to the first one). This suggests that the prompt emission
arises at a more or less constant radius and not in a propagating
single shell. This implies, in turn, that the GRB activity follows the
central engine's activity \citep{1997ApJ...485..270S} and the GRB will
last as long as the central engine is active.  Therefore, within the Collapsar
model,  the {observed} GRB duration (usually denoted
by $T_{90}$) is the difference between the engine operation time,
$t_e$, and the breakout time, $\tb$, namely $T_{90} = t_e-\tb$ (not
accounting for the redshift).  This leaves a distinctive mark on the
duration distribution of Collapsars: if $\tb$ is not negligible
compared with the typical burst duration, the observed duration
distribution of LGRBs will have a plateau at durations that are short
compared to the typical $\tb$ corrected for the typical redshift
\citep[][hereafter BNPS12,
13]{2012ApJ...749..110B,2013ApJ...764..179B}.

Such a plateau indeed appears in the duration distributions of LGRBs
observed by all three major GRB satellites: {\it Swift}, BATSE and
{\it Fermi}, with a typical breakout time of $\tb\sim10-15\;$ s (BNPS12).
These observations support both the Collapsar model and the conclusion
that the distribution of $T_{90}$ reflects both the activity time of
the central engine and the breakout time of the jet.  Following this
result we adopt this interpretation in the rest of the paper.
Interestingly, the breakout time {inferred from observations} agrees
well with the expected breakout time of hydrodynamic jets (BNPS11).

While the observations are consistent with the expected breakout times
of hydrodynamic jets, most current jet launching models are based, in
one way or another, on a Poynting flux dominated outflow.  In AGNs,
this is the only viable option. While in GRBs thermally driven
hydrodynamic jets (fireballs) are also possible, it is generally
expected that those will be less powerful than the accompanying
electromagnetic jets
\citep[e.g.][]{2013ApJ...766...31K}.  Motivated by these
considerations we (BGLP14) have recently investigated the propagation
of a magnetic jet through a stellar envelope. Unlike hydrodynamic jets
that typically cross the star at sub-relativistic velocities, Poynting
flux dominated jets have a narrower head and therefore encounter less
{resistance} by the stellar material and they cross the star at almost the
speed of light. This results in a much shorter breakout time, which
would lead to a different GRB duration distribution.

We consider here the implications of these recent findings on the type
of jet that is expected to propagate in the star.  We begin, in
\S~\ref{sec:obs} with a brief review {of} the evidence for a
plateau in the observed duration distribution of the different GRB
satellites, updating the analysis to include the recent data from {\it
Swift} and {\it Fermi}.  In \S~\ref{sec:Propagation} we summarize the
analytic results concerning the propagation of hydrodynamic and
Poynting flux dominated jets.  In \S~\ref{sec:implications} we examine
the implications of the  breakout time
{inferred from observations} on nature of the jet. We consider
the implications to three popular engines that may power the jet: a
rapidly rotating accreting black hole (BH;
\citealt{1977MNRAS.179..433B}), a BH accretion disk
\citep{1982MNRAS.199..883B, 2000ApJ...541L..21U},
and a proto-magnetar \citep{1992Natur.357..472U,2007ApJ...659..561M}. 
{Our conclusions} concerning the  magnetization of Collapsar 
jets {are summarized} in \S~\ref{sec:conclusions}, where we also
mention possible caveats to {these conclusions}.  Finally, we
discuss the implications for the composition of the jets that emerge
from the stellar envelope and produce the GRB.
 

\section{Observations}  \label{sec:obs}

GRBs are traditionally divided into two groups, long and short.
Following the original observations by BATSE
\citep{1993ApJ...413L.101K} the {dividing} line {in} 
duration is usually placed at an observed duration, $T_{90}$, of
$2\;$s.  Naturally, there is an overlap between the {two}
groups, especially at short durations
\citep{2007PhR...442..166N,2013ApJ...764..179B}.  Although the 
high-energy properties of LGRBs and short GRBs (SGRBs) are rather
similar, it was realized early on that the two groups have different
progenitors: i) The prompt high energy spectrum of LGRBs is softer on
average than that of SGRBs
\citep[e.g.][]{1993ApJ...413L.101K,2007PhR...442..166N,2013ApJ...764..179B}.
We will use this feature in the following to distinguish between the
two populations.  ii) The observed redshift distribution of SGRBs is
different \citep{1995ApJ...444L..25C}, with SGRBs typically at lower
redshifts.  iii) Their environments are significantly different: LGRBs
are observed exclusively in star forming galaxies and are associated
with the most active star forming regions within these galaxies
\citep{2006Natur.441..463F}. SGRBs, on the other hand, are observed in
a wide variety of galaxy types and within regions with different star
formation rates
\citep[e.g. see reviews by][]{2007PhR...442..166N,2013arXiv1311.2603B}. 
iv) Finally, while some low redshift LGRBs (and {\it ll}GRBs) were
observed associated with SNe, no SGRB was observed with such {an}
association.

\begin{figure}[h]
  \centering
  \includegraphics[width=12cm]{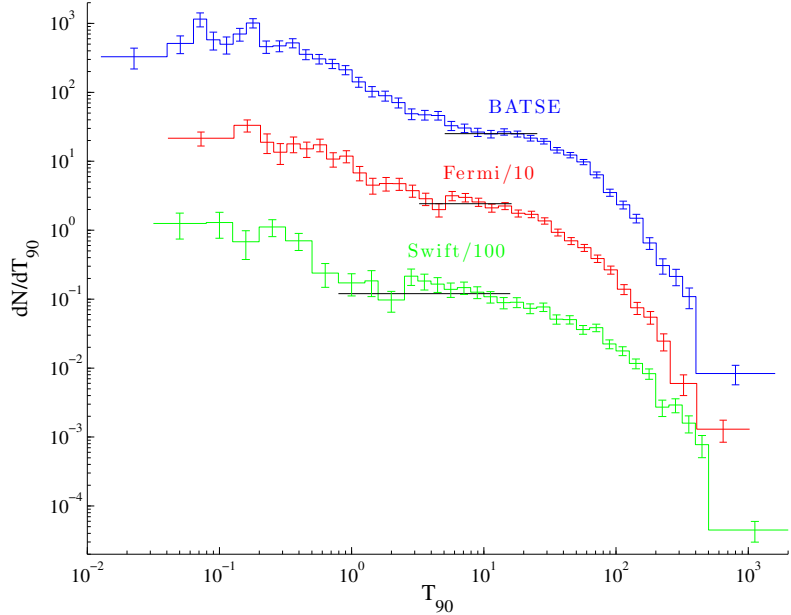}
  \vspace{-0.5 cm}
  \caption{The duration distribution, $dN_{\rm GRB}/dT_{90}$ of BATSE (blue), 
  {\it Fermi} (red) and {\it Swift}  (Green) GRBs. 
  The different curves are shifted so that they won't overlap each other. 
  The {data bins} are evenly spaced in logarithmic scale 
   with $\Delta\log(T_{90})=0.1$. Bins with less than 5 events are combined with 
   their neighbor for statistics significance. The black horizontal lines mark the bins 
   that fit a plateau at a confidence interval up to $2 \sigma$.}
  \label{fig:Plateau_all}
\end{figure} 

BNPS12 and BNPS13 {have} studied the the distribution of GRB
durations, $dN_{\rm GRB}/dT_{90}$.  As the number of GRBs observed by
{\it Swift} and {\it Fermi}-GBM has increased significantly, we update
these results, following the same methodology, using the most recent
data from these two satellites.  Figure \ref{fig:Plateau_all} depicts
the duration distribution, $dN_{\rm GRB}/dT_{90}$, of
BATSE\footnote{http://gammaray.msfc.nasa.gov/batse/grb/catalog/current/
from April 21, 1991 until August 17, 2000.}  (2100 GRBs), {\it
Fermi}-GBM\footnote{http://heasarc.gsfc.nasa.gov/W3Browse/fermi/fermigbrst.html,
from July 17, 2008 until February 14, 2014.} (1310 GRBs) and {\it
Swift} \footnote{http://{swift}.gsfc.nasa.gov/archive/grb\_table/, 
from December 17, 2004 until
February 14, 2014.} (800 GRBs). To fit a plateau in each data set we
looked for the maximal number of bins that are consistent with a
plateau at a confidence level {$\leq95\%$} ($2\sigma$)
\footnote{The confidence
level is defined here as $\int_{_0}^{^{\chi^2}}P(x,\nu)dx$, where
$P(\chi^2,\nu)$ is the density function of $\chi^2$ with $\nu$ degrees
of freedom \citep{1992nrca.book.....P}.}. The best fitted plateaus
extend from $5-25\;$s in the BATSE data 
(7.19/4 $\chi^2/DOF$), from $2.5-17\;$s in 
the {\it Fermi}-GBM
data (10/5 $\chi^2/DOF$) and from
$1-20\;$s in the {\it Swift} data (15.85/9 $\chi^2/DOF$).  We account for three free parameters in
our fit: the hight of the plateau and the two opposite ends of the
plateau line.  The differences between the maximal durations of the
plateaus can be mostly attributed to the different sensitivity and
triggering algorithms of the different detectors.

\begin{figure}[h]
  \centering
  \includegraphics[width=12cm]{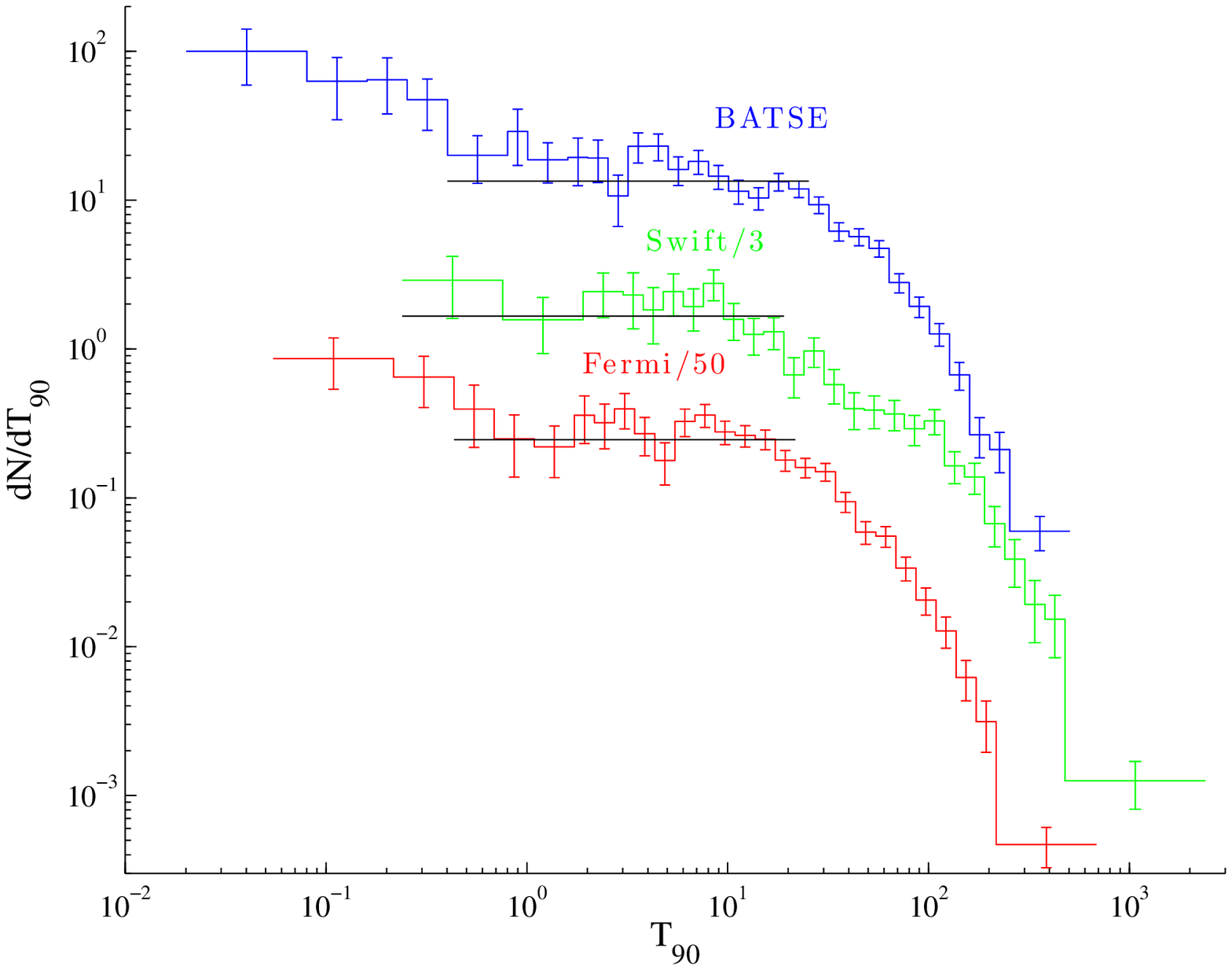}
  \vspace{-0.5 cm}
  \caption{The duration distribution, $dN_{\rm GRB}/dT_{90}$ of the soft GRBs.
  The analysis is the same as in fig \ref{fig:Plateau_all}, only the data from each satellite 
  contain only events that are softer than the median hardness 
  of the LGRBs with durations $T_{90}>20\;$s. For the BATASE this corresponds to GRBs 
  having a hardness ratio $HR_{32}<2.6$. 
  For {\it Fermi} the GRBs have a powerlaw spectral index $<-1.5$  and for
  {\it Swift}  GRBs the spectral index $<-1.7$. The analysis here
  updates the analysis in \citet{2013ApJ...764..179B} using the newer data.}
  \label{fig:Plateau_soft}
\end{figure}

At short durations the plateau is concealed by the increasing number
of non-Collapsar (``short") GRBs having a typical duration $\simless$
a few seconds
(BNPS13).  As non-Collapsars have, on average, harder spectrum than
Collapsars \citep[e.g.][]{1993ApJ...413L.101K} we can reduce the
relative number of non-Collapsars by choosing a hardness threshold
(for each sample) and selecting only the events that are softer than
this threshold.  This should lead to a less prominent ``bump" at short
duration. If the plateau is indeed an intrinsic property of the
(softer) Collapsars duration distribution, it should extend to shorter
durations in a softer subsample.  To examine this effect we select
in each sample all the events that are softer than the median hardness
of LGRBs ($T_{90}>20\;$s) in the sample (see BNPS13 for further
details).  Fig. \ref{fig:Plateau_soft} shows the duration distribution
of the soft GRB subsamples. The plateaus indeed extend to much shorter
durations than in the complete samples,
{supporting} our hypothesis.  The best fitted plateaus extend
from $0.4-25\;$s in the BATSE data (20.75/12 $\chi^2/DOF$), 
from $0.4-17\;$s in the {\it Fermi}-GBM
data (8.7/10 $\chi^2/DOF$) and from
$0.2-20\;$s in the {\it Swift} data (9.04/8 $\chi^2/DOF$).

Taking a median redshift of $z\simeq 2$ for {\it Swift} GRBs and
$z\simeq1$ for {\it Fermi} and BATSE bursts we find that in the GRBs'
{cosmological} frame these plateaus extend to $7-12\;$s,
consistent with the results obtained by BNPS12.  Note that the actual
$\tb$ may be somewhat longer than the duration that marks the end of
the plateau, but it cannot be shorter. We use the duration interval of
$7-12\;$s as our best estimate for the typical $\tb$.

A second quantity that plays an important role in our analysis is the
luminosity of the jet.  The isotropic equivalent energy of LGRBs spans
several orders of magnitudes in range, from $\sim10^{49}\;$erg to
several times $10^{54}\;$erg.  The true energy, however, is more
narrowly distributed. \citet{2003ApJ...594..674B} estimated the real
energy to be $\sim1.33\times10^{51}\;$erg, and a mean jet opening
angle of $\theta_j\simeq7^\circ$. Later works found this distribution
to be wider, with a larger tail toward lower energies than toward
higher energies.  Taking a typical LGRB duration of $\sim30\;$s this
energy corresponds to a typical one-sided jet luminosity of
$L_j\sim2\times10^{49}\;$erg/s. \citet{2005ApJ...619..412G} found
similar values of typical jet luminosity and opening angle. Estimates
of the true jet power show that a significant fraction of the jet
power is emitted as $\gamma$-rays, during the prompt phase
\citep{2001ApJ...560L..49P}.  BNPS11 have shown that in a hydrodynamic
jet the opening angle that is measured during the afterglow phase, is
very close to the injection angle of the jet at the source (see
however, \citealt{2013ApJ...777..162M} who claim that the injection
angle is a few times larger than the observed one).  In the following
we will use the observed $L_j$ as a proxy for the true jet power
during its propagation within the star adopting a canonical value of
$2\times10^{49}\;$erg/s. For hydrodynamic jets we will use the opening
angle of $7^\circ$ as a canonical value for the injection angle.
These jet properties correspond to an isotropic equivalence luminosity
of $\sim5\times10^{51}\;$erg/s, consistent with the peak flux
distribution found by \citet{2010MNRAS.406.1944W}.


\section{The breakout time of a hydrodynamic and a magnetic jet}
\label{sec:Propagation}
 
As long as the jet's head is within the star it pushes the stellar
material in front of it, forming a bow shock ahead of the jet and a
cocoon of the shocked stellar material around it. The cocoon applies
pressure on the jet and collimates it, thus changing its propagation
velocity.  The dynamics of this jet-cocoon system depends, among other
things, on the magnetization of the jet.  
Unlike a hydrodynamic jet, a highly magnetized jet cannot
effectively decelerate by shocks. Therefore, in order to match the
speed of the shocked external medium at its head, it instead gradually
decelerates by becoming narrower toward its head. This results in a
 head with a smaller cross section, 
that propagates faster through the star than its
hydrodynamic counterpart (BGLP14).

The jet's energy is dissipated at the head of the jet and flows into
the cocoon.  To continue propagating the head depends on the supply of
fresh energy from the source. If the engine stops injecting energy,
the head will essentially stop propagating once the information about
the energy cutoff will reach it.  The breakout time, $\tb$, is defined
as the time of the engine shutoff for which the information about the
shutoff reaches the jet's head when it is at the edge of the star.  If
the engine stops working at a time {$t_e$}$<\tb$, the head
will ``feel" this while it is in the star and will stop propagating.
In this case the jet will not break out and it will not produce a regular
GRB\footnote{A failed jet produces, most likely, a {\it ll}GRB when a shock wave
generated by the the dissipated energy breaks out from the seller envelope.}. Since the information
travels outwards at roughly the speed of light, the breakout time is
related to the time at which the jet's head reaches the edges of the
star as:
\begin{equation}\label{eq:t_b}
  \tb=\int_{_0}^{^{R_*}} \frac{dr}{\bh(r)c}-\frac{R_*}{c}\equiv
  \frac{R_*}{c}\frac{1-\bhbar}{\bhbar},
\end{equation}
where $\bh(r) c$ is the instantaneous   jet head velocity at $r$, and
$\bhbar c$ is the average  velocity.

Following BNPS11 and BGLP14 we obtain 
approximate analytic solutions to Eq. \ref{eq:t_b}  for   the  non-relativistic 
and the relativistic limits. 
The former is characterize by 
$\Gh\bh\ll1$, where $\Gh$ is the corresponding Lorentz factor and in this limit $\tb\simeq R_*/\bhbar c$. The latter is characterized by 
$\Gh\bh\gg1$ and in this case $\tb\simeq R_*/2\Ghbar^2c$.
The transition between the {two} limits 
occurs when $\tb\simeq R_*/c$, which according to Eq. \ref{eq:t_b} corresponds
to $\bhbar\simeq1/2$. In  the steep density profile of the stellar interior
the jet's head, which is initially sub-relativistic,
accelerates. Therefore, if the jet becomes relativistic at some
{radius}, $\rrel$, where $\Gh\bh\simeq1$, {then} it
will remain so until {it will} break out.

For a hydrodynamic jet, $\tb$ in the non-relativistic
limit is obtained by integrating $\bh$  using Eq. B3 in BNPS11: 
\begin{equation}\label{eq:t_th_hydro}
t^{^{NR}}_{\rm b,hyd}\simeq 37~L_{48}^{\dash1/3}
R_{*,4R_\odot}^{2/3}M_{*,15M_\odot}^{1/3} \theta_{0.84}^{4/3}
\fracb{3-\xi}{0.5}^{7/15}\fracb{5-\xi}{2.5}^{4/15}~{\rm s}.
\end{equation}
As canonical parameters  we have use here a stellar mass of $M_*=15 M_\odot$, a stellar
radius $R_*=4R_\odot$ 
and we assume a power-law density profile: $\rho_*\propto r^{-\xi}$
with $\xi=2.5$.  Hereafter we measure masses and radii in units of
solar mass and solar radius respectively and use the subscript `$*$'
to denote properties of the progenitor star. For all other quantities
we use the dimensionless form $A_x\equiv A/10^x$ {measured in c.g.s
units.}  In the relativistic limit $\tb$ is obtained by approximating
$\bh\simeq1-1/2\Gh^2$, and using Eq. B14 from BNPS11:
\begin{equation}\label{eq:t_th_rel_hydro} 
  t^{^{R}}_{\rm b,hyd}\simeq 2~
L_{52}^{-1/5}R_{*,4R_\odot}^{4/5}M_{*,15M_\odot}^{1/5}\theta_{0.84}^{4/5}
\fracb{3-\xi}{0.5}^{7/25}\fracb{5-\xi}{2.5}^{4/25}\fracb{4.5}{7-\xi}~{\rm
s}.  
\end{equation}  

\begin{figure}[h]
   \centering
  \includegraphics[width=12cm]{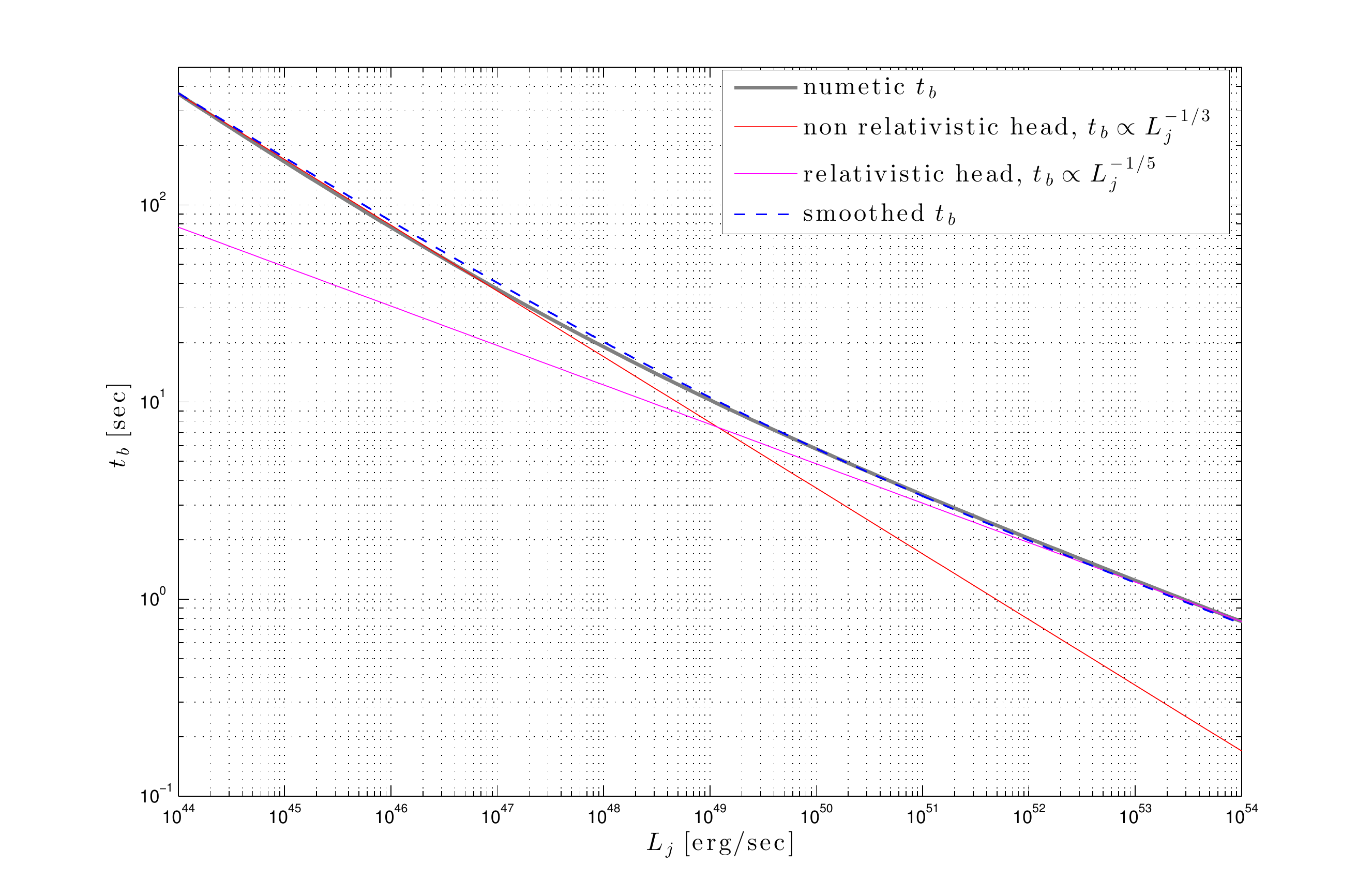} 
  \vspace{-0.5cm}
  \caption{The breakout time, $\tb$, as a function of $\Lj$ calculated 
  for a jet with an opening angle $\theta_j=7^\circ$, and a star with a mass
  $M_*=15M_\odot$, radius $R_*=4R_\odot$ and a power law density profile 
  $\rho\propto r^{-2.5}$. The gray solid curve tracks the exact integration of Eq. \ref{eq:t_b},
  the red and magenta lines show the analytic approximation for the non relativistic
  and the relativistic cases respectively. 
  The dashed blue line follows the smoothed analytic solution for $\tb$
  from Eq.\ref{eq:t_b_smooth}.}
 \label{fig:tb_ann}
\end{figure}

Figure \ref{fig:tb_ann} depicts $\tb$ as a function of $\Lj$ calculated
with the fiducial values of injection angle and stellar properties.  
The gray line tracks the exact integration of
Eq. \ref{eq:t_b}.  The solid red and magenta lines show the analytic
approximations at the low luminosity (non-relativistic) asymptote and
at the high luminosity (relativistic) asymptote respectively.  The
transition between the asymptotes occurs at a luminosity range of
about $10^{48}-10^{51}$ ergs/s (for our chosen set of parameters),
which is the range that is relevant for collapsar jets. 
To obtain a useful analytic solution we approximate the exact
integration as (dashed blue line in Fig. \ref{fig:tb_ann}):
\begin{equation}\label{eq:t_b_smooth}
  t_{\rm b, hyd}\simeq6.5R_{*,4R_\odot}
  \left[\fracb{\Lj}{L_{\rm rel}}^{-2/3}+\fracb{\Lj}{L_{\rm rel}}^{-2/5}\right]^{1/2},
\end{equation}
where $L_{\rm rel}$ is the transition luminosity between a non relativistic
breakout time and a relativistic one:
\begin{equation}\label{eq:L_switch} 
 L_{\rm rel}\simeq 1.6\times10^{49}~
R_{*,4R_\odot}^{-1}M_{*,15M_\odot}\theta_{0.84}^{4}
\fracb{3-\xi}{0.5}^{7/5}\fracb{5-\xi}{2.5}^{4/5}\fracb{7-\xi}{4.5}^{15/2}~{\rm erg/s}\ .  
\end{equation} 
{The corresponding breakout time is } $\tb(L_{\rm rel})\simeq9
R_{*,4R_\odot}\;{\rm s} \simeq R_*/c$, as expected.  A typical
collapsar jet with luminosity of $\sim2\times10^{49}$ erg/s and
$\theta \sim 7^\circ$, is therefore at most mildly relativistic by the
time it breaks out of the star.

A Poynting flux dominated jet becomes relativistic deep inside the
star, even with a modest power (BGLP14):
\begin{equation}\label{r_1_mag} \frac{\rrel}{R_*}\simeq
1.4\times10^{-2} 
\left[L_{49.3}^{-1}M_{*,15M_\odot}R_{*,4R_\odot}^{-3}r_{_{L,7}}^2
\fracb{3-\xi}{0.5} \right]^{1/\xi}.
\end{equation}
This  implies that here only the relativistic asymptotic solution is relevant.
The corresponding breakout time is obtained, accordingly, using the
relativistic approximation of $\bh$ in BGLP14:
\begin{equation}\label{eq:t_th_mag}
  t_{\rm b,mag}\simeq
  0.8~L_{49.3}^{-1/3}M_{*,15M_\odot}^{1/3}r_{_{L,7}}^{2/3}\fracb{0.5}{3-\xi}^{2/3}  ~ {\rm s}.
\end{equation}
This time is much shorter than the breakout time 
of a hydrodynamic jet with a similar luminosity. 


\section{Implications} \label{sec:implications}
The observed GRBs duration distribution indicates a typical breakout
time of $\sim7-12\;$s.   This timescale arises naturally for
hydrodynamic jets (Eq. \ref{eq:t_b_smooth}) and it is much longer than typical breakout
times of Poynting flux dominated jets (Eq.  \ref{eq:t_th_mag}). 
For the canonical hydrodynamic jet  used in this work  these breakout times imply a stellar radius of  $R_*\simeq\left(3 M_{*,15M_\odot}^{-1/2}-
 6 M_{*,15M_\odot}^{-1/4}\right)R_\odot$,  which is consistent with expected radii of
Wolf Rayet stars that are the likely progenitors of LGRBs.
The different  {power-law indices} in the 
mass arise  from the fact that
{for} $3R_\odot$ the jet's head is non-relativistic ($L_{\rm rel}<\Lj$)
while {for}  $6R_\odot$ it is relativistic   ($\Lj>L_{\rm rel}$, see Eq. \ref{eq:t_b_smooth}).

For a Poynting dominated jet these breakout times could arise only
with a light cylinder radius of
$\rL\sim\left(2.5-5\right)\times10^8\;$cm (Eq. \ref{eq:t_th_mag}).
This value is larger than the one that expected for GRBs central
engines, thus disfavoring this option.
Before ruling out Poynting flux dominate jets we turn to examine three
central engine models and explore whether such a large value of the
light cylinder is a viable option. Clearly, 
unlike the general discussion we had so far,  this discussion is
model dependent, as assumptions have to be made on the specific nature of the central engine and the jet acceleration mechanism. Still it gives a good
indication on the question {of} whether a 
central engine producing such a Poynting flux dominated jet is plausible.

The light cylinder radius, $\rL\sim(2.5-5)\times10^8$ cm, corresponds to
an angular frequency of the magnetosphere at the base of the jet {of}
$\Omega_m=c/\rL\simeq60-120\;$rad/s.  It is generally accepted that
the Poynting dominated jet is driven by the rotation of the central
engine.  The field lines that are connected to the rotating central
object are winded {up} by the rotation, generating a toroidal field. The
toroidal ``hoops" then propagate outward under their own pressure and
transfer most of the energy outward in the form of Poynting
flux. Therefore, the implied value of $\Omega_m$ has a direct
consequence on the rotation and on the luminosity that can be
generated by the central object. We consider the implications on three
objects that have been suggested to power a Poynting dominated jet: An
accreting, rotating BH
\citep[e.g.][]{1977MNRAS.179..433B}, a magnetized accretion disk
\citep[e.g.][]{1982MNRAS.199..883B,2000ApJ...541L..21U, 2002ApJ...572..445L}
and a rapidly rotating magnetar \citep{1992Natur.357..472U,2007ApJ...659..561M}.

In an accreting rotating BH, the magnetic field lines of the jet
thread the horizon of the BH.  The horizon has an {effective}
resistance of $4\pi/c$, which produces a drag on the field lines and
causes them to rotate at $\Omega_m\simeq \eta\Omega_H$, where
$\Omega_H$ is the angular velocity of the BH and $\eta$ is usually
estimated as $1/2$
\citep[e.g.][]{1977MNRAS.179..433B}.   
The BH's angular velocity  satisfies $\Omega_H=ac/R_H$, 
where $a$ is the dimensionless 
spin of the BH and $R_H=GM_H\left(1+\sqrt{1-a^2}\right)/c^2$ is the radius of the
horizon giving:
\begin{equation}
  \frac{a}{1+\sqrt{1-a^2}}\simeq5\times10^{-3}\fracb{\eta}{0.5}^{-1}\Omega_{m,2}
  M_{H,5M_{\odot}}.
\end{equation}
The limit  $\Omega_m \simeq60-120\;$rad/s implies a very low value of 
$a\approx(3-6)\times10^{-3}$. The corresponding power output that goes into
to each jet is 
\citep{1977MNRAS.179..433B}:
\begin{equation}\label{eq:L_j_BH}
  L_j\simeq5.5\times10^{49}B_{16}^2M_{5M_\odot}^2
  \fracb{a+a\sqrt{1-a^2}}{0.02} ^2{\rm erg/s}\ .
\end{equation}
Thus the slow rotating BH can provide the observed luminosity of LGRBs, 
if the magnetic field on the horizon is about $\sim10^{16}\;$G.
Such a high magnetic field requires an efficient amplification process 
either in an accretion disk or in the core, prior to its collapse, 
and is likely accompanied by a significant rotation.
It is hard to imagine how the BH can acquire such a high magnetic field
without being spun up by the angular momentum of the accreted matter,
making this scenario less realistic.

The jet might be launched from the accretion disk itself.  A
magnetized accretion disk rotates differentially around the BH.
Magnetic field lines that are connected to the disk at different radii
rotate at different angular velocities and have different light
cylinder radii.  Therefore if the jet is powered by the accretion
disk, the constraint on $\Omega_m$ of the jet implies a constraint on
the disk's radius where most of the jet's energy is injected.  Assuming
a Keplerian (and for simplicity Newtonian) disk, this radius is: $
\simeq 50~r_{_{\rm g}}~\Omega_{m,2}^{-2/3}M_{H,5M_\odot}^{-2/3}$,
where $r_{_{\rm g}}=GM_H/c^2$ is the gravitational radius of the BH.
Such an injection radius
seems too large for  realistic engine disk
models. For example, 
\citet{2002ApJ...572..445L,2003ApJ...596L.159L} found that a
Keplerian disk around a Schwarzschild BH radiates almost all of its
Poynting flux inside it's last stable orbit radius, located at $\sim 6r_g$. 
Note that a coupling
in the radial direction between different parts of the disk will only
increase the rotational velocity at each radius and will further
increase {the radius corresponding to a given $\Omega_m$}.

In a third scenario the jet is powered by a rapidly rotating
magnetar. In this case the magnetic field lines are anchored to the
surface of the NS and corotate with it. Therefore the bounds we infer
for $\Omega_m$ determine the rotational velocity of the magnetar.  The
power that goes into the jet can be evaluated by integrating the
Poynting flux, $\frac{c}{4\pi}{\bf E}\times{\bf B}$, across the field
lines that are associated with the jet. The power is maximized when we
assume that all the field lines are channeled into the jet and the
magnetosphere has the topology of a split monopole near the base of
the magnetar. {Such a topology is expected, at least in the 
early stages of the magnetar, when the magnetosphere is still 
baryon loaded by the neutrino-driven wind  blowing from the 
surface of the magnetar \citep{1992Natur.357..472U, 2007ApJ...659..561M}}.
This gives a one-sided jet power of: 
\begin{equation}\label{eq:NS_jet_power}
 \dot{E}\simeq 10^{49}B_{16}^2 R_{6}^4\Omega_{2}^{2} ~{\rm erg/s},
\end{equation}
{where $B$ is the magnetic field at the surface of the magnetar. 
A configuration of a force free magnetosphere will give {an even lower}
luminosity {than} that \citep{2006ApJ...648L..51S}.
It can be seen that a typical power of LGRB jets can barely be
satisfied with the inferred rotational velocity, {even in the extreme
condition where all magnetic field lines are open, and} only if
{$B\sim10^{16}R_6^{-2}\;$G}.  Note that
\cite{2012MNRAS.422.2878D} suggest that even though magnetars are born
with such {internal} fields, their {initial} dipole
field that is relevant here is $\lesssim 10^{15}\;$G. Like in the case
of the BH it is hard to see how such high field can be obtained
without {requiring a} significant
rotation of the magnetar.  Moreover, estimates of the rotational
energy of the magnetar, assuming it rotates rigidly, give
\begin{equation}\label{eq:NS_Erot}
  E_{\rm rot}\approx 5\times 10^{48}\Omega_{2}^{2}R_{6}^{2}
  M_{1.4\rm M_\odot}, 
\end{equation} 
where $R$ and $M$ are the radius and mass of the magnetar.  This
rotational energy is about two orders of magnitude lower than the
energy of a typical LGRB. Thus the magnetar scenario is also rejected.

We find that in all three central engine models that we considered: an
accreting BH, a NH accretion disk or a rapidly rotating magnetar, are 
incompatible with a large $\rL$ that is required to produce the  
breakout times of $7-12$ s, inferred from the duration distribution. 
This {implies} that the jet must be launched
hydrodynamically or alternatively if the jet  is launched 
Poynting flux dominated  it should dissipate its magnetic energy
rapidly within the star and become hydrodynamic. 
In either case the jet emerging 
from the stellar envelope is not Poynting flux dominated.


\section{Conclusions}\label{sec:conclusions}

The lack of a strong evolution in the properties of the prompt
emission of LGRBs suggests that the conditions within the emission
regions are rather constant in time. A single moving shell would have
expanded by a factor of $\sim 10-100$ during the duration of a burst
and it is unlikely to maintain constant conditions as it emits the
prompt gamma-ray emission over such a wide range of radii.  This, in
turn, suggests that the duration of the burst is determined by the
activity of the central engine and not by a local process within the
emission region.  In a Collapsar model in which the jet has to break
out from the progenitor star we therefore expect that $T_{90} = t_e -
\tb$. This relation implies a plateau in the duration distribution of
GRBs at durations shorter than the typical breakout time. Indeed, such
a plateau exists in the duration distribution of the three major GRB
satellites, {\it Swift}, {\it Fermi} and BATSE at durations shorter than
$\sim 25\;$s \citep{2012ApJ...749..110B}, confirming the basic assertions of this model.
Reanalysis of the latest {\it Swift} and {\it Fermi} data confirms the
presence of this plateau up to observed durations of $\sim 17-25\;$s,
which correspond to a breakout time $\tb\sim7-12\;$s in the source
frame. 

The inferred breakout time is consistent
with the propagation time of a hydrodynamic jet, and it implies a
progenitor radius of $\sim(3-6)R_\odot$.  On the other hand, this
breakout time is too long for  typical parameters expected for
a Poynting dominated jet.  A breakout time of $\sim7-12\;$s requires a
light cylinder radius of $\rL\simeq(2.5-5)\times10^8\;$cm, corresponding
to an angular frequency of $\Omega_m\simeq60-120\;$rad/s at the base
of the jet. 

Such a large value of $\rL$ and the corresponding low $\Omega_m$ are
inconsistent with the three most popular Poynting flux jet
acceleration models.  If the jet is powered by an accreting BH, this
angular frequency implies a very low {spin parameter} of the BH,
$a\simless0.01$.  This value is most likely too low to allow an
amplification of the magnetic field to $\sim10^{16}\;$G, required to
power the GRB jet. {Similarly,} the low $\Omega_m$ renders magnetars
as an unlikely sources. The magnetar would simply rotate too slow to
supply the observed power and the total energy.  Finally, an accretion
disk is unlikely to power the jet as well, since the low rotation rate
implies that it should inject most of the energy at a distance of
about 50 gravitational radii, too far for reasonable disk engine
models.

We conclude that during most of its propagation within the star the
jet has a low magnetization and it propagates as a hydrodynamic jet.
This result leads to some interesting implications on the properties
of LGRB internal engines and the conditions at the base of the
jets. One possibility is that the jet is launched hydrodynamically at
the source. The most probable process for that is neutrino
annihilation above the rotational axis of the central engine
\citep[e.g.][]{1989Natur.340..126E,1993ApJ...418..386L}. 
This scenario  can work
only if the accretion rate is $\gtrsim 0.1\,M_\odot$/s, so that 
neutrino emission is large enough to power the observed jets
\citep{2013ApJ...766...31K,2013ApJ...770..159L}. 
The high accretion rate must be sustained throughout the entire 
duration of the GRB which can last from tens to hundreds seconds.
Though a duration of $\lesssim 30\;$s seems to be consistent 
with such a model  \citep[e.g.][]{2010ApJ...713..800L}, its seems unlikely
to be capable of powering longer duration GRBs.

A second possibility is that the jet is launched Poynting dominated
{but} it dissipates {most of} its magnetic energy close to the
source, {and} it then propagates as a hydrodynamic jet. An appealing process
for such efficient dissipation is the kink instability
\citep{1992SvAL...18..356L, 1993ApJ...419..111E, 1997MNRAS.288..333S, 
1998ApJ...493..291B, 1999MNRAS.308.1006L, 2006A&A...450..887G}.
However, we (BGLP14) have shown that Collapsar jets are unlikely to be
disrupted by {the} kink instability. Thus, the dissipation
should arise from a different process.

A third possibility is that the jet changes its character with time.
Our conclusion concerning the jet composition applies only to the
initial phase, while its head is still within the stellar
envelope. This phase, which lasts about 10 seconds, must be
predominantly hydrodynamic. Once the jet has breached out of the star
it could be Poynting flux dominated. This would require, of course, a
more complicated central engine that switches from one mode to
another. While this seems contrived, remarkably, some magnetar models
suggest such a possibility
\citep{2011MNRAS.413.2031M}. 
One can also imagine accretion disk models that initially cool via neutrinos 
and later on as the accretion rate decreases, becomes Poynting flux dominated 
\citep{2013ApJ...766...31K}. However, all such models require 
some {degree of coincidence} as 
the central engine doesn't receive any feedback form the propagating
jet and there is no a priori reason that the transition from one
composition to the other would take  place  {just} at the right
stage.

Before concluding we note that our interpretation of the observed
temporal duration distribution depends on the condition
$T_{90}=t_e-\tb$.  This condition holds if the prompt $\gamma$-rays
activity follows the activity of the central engine and if the jet has
to penetrate the envelope of the collapsing star.  The first
assumption is supported by observational features of the prompt GRBs.
The second one is the central assumption of any Collapsar model.  With
both assumptions and with $\tb \sim 10$ s the plateau and the break
{in the duration distribution} arise naturally for Collapsars with
hydrodynamic jets.  If one assumes that $T_{90}$ is not determined by
the central engine or if one asserts that $\tb \ll T_{90}$ (in which
case the plateau {does not} reflect the jet breakout time),
one has to suggest a model for the GRB duration distribution that
explains the observed plateau at short durations and the break at
{$\sim 10\;$s}, as well as other temporal properties of LGRBs.  We are not {aware} of such a model.

We conclude that regardless of how the jets are formed, the jets that emerge from the Collapsar envelopes are not Poynting
flux dominated.  This result is consistent with the recent findings of
\citet{2014arXiv1402.4113B}. These authors have shown, based on the
properties of the observed spectra, that it is unlikely that the
emitting region, where the prompt $\gamma$-rays are produced, is
Poynting flux dominated.

\subsection*{acknowledgements}
We thank E. Nakar,  A. Philippov \& A. Tchekhoskoy for meaningful discussions.
This research was supported by the ERC advanced research grant ``GRBs'' 
by the  I-CORE (grant No 1829/12) and by HUJ-USP grant and by a grant from 
the Israel space agency, SELA.


\bibliographystyle{mn2e.bst}

\end{document}